\documentclass[prd,reprint,superscriptaddress,nofootinbib,longbibliography,aps]{revtex4-1}

\usepackage{graphicx}
\usepackage{longtable}
\usepackage{amsmath}
\usepackage{color}
\usepackage{xcolor}
\usepackage{amssymb}
\usepackage{subfigure}
\usepackage{tabularx}
\usepackage[colorlinks=true]{hyperref}
\usepackage{cleveref}
\usepackage[normalem]{ulem}
\usepackage{slashed}
\usepackage[toc]{appendix}
\usepackage{bm}
\usepackage{braket}
\usepackage{comment}
\usepackage{orcidlink}

\newcommand{\hato}{\hat\Omega}

\begin{document}

\title{Analytical Estimates of Gravitational Wave Background Anisotropies from Shot Noise and Large-Scale Structure in Pulsar Timing Arrays}

\author{Meng-Xiang Lin\,\orcidlink{0000-0003-2908-4597}}
\altaffiliation{CITA National Fellow}
\email{mengxiang\_lin@sfu.ca}
\affiliation{Department of Physics, Simon Fraser University, Burnaby, British Columbia, V5A 1S6, Canada}
\affiliation{Canadian Institute for Theoretical Astrophysics (CITA), University of Toronto, 60 St George Street, Toronto, Ontario M5S 3H8, Canada}
\affiliation{Center for Particle Cosmology, Department of Physics and Astronomy, University of Pennsylvania, Philadelphia, Pennsylvania 19104, USA}

\author{Adam Lidz\,\orcidlink{0000-0002-3950-9598}}
\affiliation{Center for Particle Cosmology, Department of Physics and Astronomy, University of Pennsylvania, Philadelphia, Pennsylvania 19104, USA}

\author{Chung-Pei Ma\,\orcidlink{0000-0002-4430-102X}}
\affiliation{Department of Astronomy, University of California, Berkeley, California 94720, USA}
\affiliation{Department of Physics, University of California, Berkeley, California 94720, USA}

\begin{abstract}
An important next step for pulsar timing arrays (PTAs) is to measure anisotropies in the gravitational wave background (GWB) at $\sim$ nano-Hz frequencies. 
We calculate the expected GWB anisotropies using empirically calibrated models for the merger rates of supermassive black hole binaries (SMBHBs).  
The anisotropies reflect both shot-noise in the discrete SMBHB populations while also tracing, in part, the large-scale structure (LSS) of the universe. 
The shot-noise term is sensitive to the high-mass end of the merging SMBH mass function, depends somewhat on the low-redshift tail of the merger distribution, and is a strong function of observing frequency. The precise frequency dependence provides a test of SMBHB residence times. In our models, the mean shot-noise anisotropy typically lies close to or above the broad frequency-band NANOGrav upper limits. Consequently, near-future PTA data, and potentially re-analyses of existing measurements using frequency-dependent shot-noise anisotropy templates, should be capable of detecting this signal or placing meaningful constraints on SMBHB merger models. A full interpretation, however, will require modeling the probability distribution of shot-noise amplitudes rather than relying solely on ensemble-averaged predictions. 
The LSS-induced anisotropies are at least two to three orders of magnitude smaller. Although the LSS contribution contains valuable information regarding the redshift distribution and clustering bias of the merging SMBHBs, detecting this component will be challenging.   
\end{abstract}

\maketitle

\section{Introduction}
The strong evidence for a stochastic gravitational wave background (GWB) at nanohertz frequencies reported by Pulsar Timing Arrays (PTAs)~\cite{NANOGrav:2023gor,Xu:2023wog,EPTA:2023fyk,Reardon:2023gzh,InternationalPulsarTimingArray:2023mzf,Miles:2024seg} has opened a new observational window into both astrophysics and fundamental physics. The GWB is most plausibly sourced by inspiraling supermassive black hole binaries (SMBHBs). PTA measurements therefore provide a powerful probe of the dynamics of SMBHB mergers and the demographics of SMBH populations, especially when combined with local kinematic SMBH mass measurements and observations of distant quasars and active galactic nuclei. Together, these datasets can inform models for the formation of SMBHs and 
the co-evolution with their host galaxies. 

The GWB may also contain contributions from the early-universe, such as from cosmic inflation, topological defects, first-order phase transitions, or from primordial black holes~\cite{LISACosmologyWorkingGroup:2022kbp,NANOGrav:2023hvm}. 
Although the observed PTA signal is broadly consistent with the SMBHB scenario, there is some discussion in the literature regarding whether it matches theoretical expectations in detail.  For example, \cite{Sato-Polito:2023gym} argue that the GWB amplitude is higher than expected, while \cite{Liepold:2024woa} find better agreement with an alternate model for the local SMBH mass function. 
Ongoing pulsar monitoring efforts and developments of new analysis techniques will help to test the current understanding of SMBH populations.

One such promising direction is the study of GWB anisotropies~\cite{Mingarelli:2013dsa,Taylor:2013esa,Kramer:2013kea,Mingarelli:2017fbe,Yang:2018ycs,
Sato-Polito:2023spo,NANOGrav:2023tcn,Gardiner:2023zzr,Agazie:2024jbf,Allen:2024mtn,Raidal:2024tui,Li:2024qcs,Domcke:2025esw,Chen:2026mid,Bernardo:2025dat}.
The key to detecting the sky-averaged GWB is the Hellings-Downs curve~\cite{Hellings:1983fr}, which describes the GWB-induced correlated timing residuals between pairs of pulsars as a function of their angular separation. Anisotropies in the GWB imprint distinctive departures from, and variations around, the Hellings-Downs curve, enabling searches for or constraints on anisotropies~\cite{Mingarelli:2013dsa,Taylor:2013esa,Ali-Haimoud:2020ozu,Ali-Haimoud:2020iyz}. These searches are also intertwined with using PTA measurements for tests of modifications to general relativity, which can also lead to deviations from the Hellings-Downs curve~\cite{Lee:2010cg,Gair:2015hra,Qin:2020hfy,Liang:2021bct,Bernardo:2022rif,Liang:2023ary,Hu:2024wub,Bernardo:2024bdc}.

We expect two main contributions to the GWB anisotropies. First, the discrete nature of SMBHBs induces Poisson, or ``shot-noise'', fluctuations in the source abundance and GWB strain amplitudes across the sky. Second, the SMBHBs are expected to trace the large-scale structure (LSS) of the universe, with overdense regions in the underlying matter distribution hosting more galaxies and SMBHBs and sourcing a stronger strain signal, while underdense regions contain fewer sources and a weaker GWB signal. The shot-noise anisotropy can help quantify whether the GWB is produced mainly by a few bright sources or instead sourced by many comparably faint ones. The LSS signal depends on the redshift distribution of the sources and the clustering bias of the SMBHBs.

Our work extends earlier studies, including~\cite{Sato-Polito:2023spo}, which previously estimated the expected GWB shot-noise anisotropies. Here we update their calculations using the empirically calibrated merger rate distributions from the more recent studies of \cite{Sato-Polito:2023gym,Liepold:2024woa}. We also include analytic expressions for the shot-noise power spectrum in these models, and point out the sensitivity to the low redshift tail of the merger rate distribution, and the high mass end of the SMBH mass function. Some related discussion also appears in \cite{Sato-Polito:2024lew,Lamb:2024gbh} in the context of calculations of the probability distribution/moments of the GWB measurements versus frequency. 
The LSS signals are also considered in~\cite{Semenzato:2024mtn,Sah:2024oyg,Sah:2024etc,Cusin:2025xle,Sah:2025uuk,Sah:2026bpl}.
Here we calculate the LSS and shot-noise anisotropies using a common analytic framework and compare the
two signals.

The paper is organized as follows. 
We provide analytical formulas for both sources of anisotropy in Sec.~\ref{sec:signals} and show the results of concrete models in Sec.~\ref{sec:results}. An Appendix discusses the details of a model for the clustering bias of merging SMBHBs.  
We summarize our results and conclude in Sec.~\ref{sec:conclusions}.

\section{GWB Anisotropy Signals}\label{sec:signals}
Our aim is to estimate the GWB anisotropies from both shot-noise and the LSS, assuming that the PTA signals are dominated by the inspiral phase of SMBHBs. First, we introduce relevant definitions, and then we give general expressions for both sources of anisotropy. Finally, we model the GWB angular power spectrum from shot-noise and the LSS using concrete models for SMBHB merger rates from the current literature. 

\subsection{Angular Correlation Function and Power Spectrum of Strain Anisotropies}

In the LSS literature, 
one basic statistical quantity used to characterize galaxy inhomogeneities on the sky is
the angular two-point correlation function
\begin{equation} \label{eq:xi_gal}
    \omega_{\rm g}(\cos \vartheta) \equiv \langle \delta_{\rm g}(\boldsymbol{\hato}) \delta_{\rm g}(\boldsymbol{\hato^\prime}) \rangle \,,\quad \mathrm{cos}(\vartheta) = \boldsymbol{\hato}\cdot \boldsymbol{\hato^\prime} \,, 
\end{equation}
where $\delta_{\rm g}(\boldsymbol{\hato})$ is the fractional deviation in the number density of galaxies in direction $\boldsymbol{\hato}$, $n_{\rm g}(\boldsymbol{\hato})$, from the sky-averaged mean density $\langle n_{\rm g}\rangle$:
\begin{equation}
    \delta_{\rm g}(\boldsymbol{\hato}) = \frac{n_{\rm g}(\boldsymbol{\hato}) - \langle n_{\rm g}\rangle}{\langle n_{\rm g} \rangle} \,.
\end{equation}
By statistical isotropy, the average over the product of density contrasts depends only on the angular separation between the two points on the sky, $\vartheta$. In practice, the two-point correlation function can be estimated by considering pairs of points in various bins in angular separation.

By analogy, for the case of GWB anisotropies, we generally consider fractional fluctuations in the strain amplitudes. 
Here, the total characteristic strain, $h^2_c(f)$, is defined such that the total energy density in gravitational waves per logarithmic frequency interval, 
$d \ln f $, is $\Omega_{\rm GW} \propto f^2 h^2_c(f)$ \cite{Phinney:2001di}, after summing over all sources on the sky. Throughout, we refer to the strain-squared as a ``strain amplitude'', although this term is sometimes used in the literature to refer instead to $h_c(f)$ itself. 
We use $h^2(f,\boldsymbol{\hato})$ to denote the strain amplitude per unit steradian from sources in a direction near
$\boldsymbol{\hato}$. The angular average of this quantity, which we denote as $\langle h^2 \rangle$ for brevity (without explicitly specifying the frequency dependence) is
\begin{equation}\label{eq:h2ave}
    \langle h^2 \rangle \equiv \frac{1}{4\pi}\int d^2\hato \, h^2(\boldsymbol{\hato}) \equiv \frac{h_c^2}{4\pi}\ ,
\end{equation}
where $h_c^2$ is the dimensionless total characteristic strain while $\langle h^2 \rangle$ is in units of inverse steradian.

The fractional fluctuations in the strain amplitude are then defined, relative to the spatial average of Eq.~\eqref{eq:h2ave}, as
\begin{equation}
    \delta_{h^2}(\boldsymbol{\hato})\equiv \frac{h^2(\boldsymbol{\hato})-\langle h^2 \rangle}{\langle h^2 \rangle}\,.
\end{equation}
The corresponding angular two-point correlation function has the same form as Eq.~\eqref{eq:xi_gal}:
\begin{eqnarray}\label{eq:2pt-h2}
    \omega_{h^2}(\cos \vartheta) = \langle \delta_{h^2}(\boldsymbol{\hato}) \delta_{h^2}(\boldsymbol{\hato^\prime}) \rangle \,,\quad \mathrm{cos}(\vartheta) = \boldsymbol{\hato} \cdot \boldsymbol{\hato^\prime}\,,
\end{eqnarray}
and depends only on $\vartheta$ by statistical isotropy.

We consider two separate contributions to the two-point correlation of strain amplitude fluctuations: a shot-noise term from self-correlations and an LSS term that arises because the SMBHB mergers trace the LSS: 
\begin{equation}\label{eq:PTA-corr}
    \omega_{h^2}(\cos\vartheta) = \omega_{h^2}^{\rm LSS}(\cos\vartheta) + \frac{\langle h^4 \rangle}{\langle h^2 \rangle^2} \frac{\delta_D(1 - \cos\vartheta)}{2 \pi}\,.
\end{equation}
The first term on the right-hand side captures correlations from distinct points on the sky and is attributed to the LSS. In the second term, $\delta_D$ is a Dirac $\delta$-function in angular separation, selecting only coincident pairs of points and capturing the shot-noise self-correlations. The strength of the self-correlations depends on the fourth moment of the strain field, $\langle h^4 \rangle$ \cite{Sato-Polito:2023spo}.

The fractional strain fluctuations can be expanded in a spherical harmonic series as
\begin{equation}
    \delta_{h^2}(\boldsymbol{\hato}) \equiv \sum_{\ell,m} a_{\ell m} Y_{\ell m}(\boldsymbol{\hato})\,,
\end{equation}
where the $Y_{\ell m}(\boldsymbol{\hato})$'s are the usual spherical harmonics. 
It then follows that the angular correlation function $\omega_{h^2}$ and the angular power spectrum $C_{\ell,h^2}$ are related by
\begin{eqnarray}
\label{eq:Cl-integral}
    \omega_{h^2}(\cos \vartheta) \equiv \sum_\ell \frac{2\ell+1}{4\pi}P_\ell(\cos\vartheta) C_{\ell,h^2}\,, \nonumber\\
    C_{\ell,h^2} = 2 \pi \int_{-1}^1 d\cos\vartheta \, P_\ell(\cos\vartheta) \omega_{h^2}(\cos\vartheta)\,,
\end{eqnarray}
where $P_\ell(\cos\vartheta)$ is a Legendre polynomial and 
$C_{\ell,h^2}$ is related to
the spherical harmonic coefficients $a_{\ell,m}$ as
\begin{equation}
    C_{\ell,h^2} \equiv \frac{1}{2\ell+1}\sum_m \langle a_{\ell m}a_{\ell,m}^* \rangle\,.
\end{equation}
Similar to Eq.~\eqref{eq:PTA-corr}, both the LSS and shot-noise (SN) contribute to the angular power spectrum:
\begin{equation}
    C_{\ell>0,h^2} = C_{\ell>0,h^2}^{\rm LSS} + C_{\ell>0,h^2}^{\rm SN}\,.
\end{equation}

Note that since we work with strain amplitude contrasts, $\delta_{h^2}$, the angular power spectrum here is defined relative to the
mean strain amplitude. Our conventions differ from common ones in the GWB anisotropy literature, including
those in the recent NANOGrav anisotropy search paper ~\cite{NANOGrav:2023tcn}. In that work, the authors consider the angular power spectrum of the strain amplitude itself, rather than the fluctuations relative to the mean. The relation between the angular power spectra in the two studies is:
\begin{equation}\label{eq:cl_nanograv}
    (\tilde{C}_{\ell>0}/\tilde{C}_{\ell=0})^{\rm NANOGrav} = \frac{1}{4\pi}C_{\ell>0,h^2}\,,
\end{equation}
where the left-hand side is the quantity used by NANOGrav, while the right-hand side is the corresponding quantity according to our notation and conventions.  

\subsection{Anisotropies from Shot Noise}\label{sec:shot-noise}
The amplitude of the shot-noise signal can be characterized by an effective number of sources (per steradian) contributing to the GWB:
\begin{equation}\label{eq:Neff}
    \frac{4\pi}{N_{\rm eff}} \equiv \frac{\langle h^4 \rangle}{ \langle h^2 \rangle^2}\,.
\end{equation}
In the limit that each source has the same strain amplitude, $N_{\rm eff}$ is simply the number of sources. 
Combining Eqs.~\eqref{eq:PTA-corr} and \eqref{eq:Cl-integral} and setting $P_\ell(\cos\theta=1)=1$ for all $\ell$, we obtain an $\ell$-independent angular power spectrum for the shot-noise-induced anisotropies:
\begin{equation}
\label{eq:clsn}
    C_{\ell>0,h^2}^{\rm SN} = \frac{\langle h^4 \rangle}{ \langle h^2 \rangle^2} \equiv \frac{4\pi}{N_{\rm eff}}\,.
\end{equation}
All the terms, however, have non-trivial frequency dependence, which we examine below.

To derive an expression for $\langle h^4 \rangle$ due to a population of inspiraling SMBHBs, we recall that the
sky-averaged strain of the GWB produced by the same sources, $\langle h^2\rangle=h_c^2/4\pi$ (see Eq.~\eqref{eq:h2ave}), can be written as \cite{Phinney:2001di}
\begin{equation}\label{eq:h2c}
    \langle h^2\rangle(f)= 
    \frac{G}{\pi^2 c^2f^2}\int 
    d\mathcal{M}\frac{dz}{1+z} \frac{d^2 n}{d\mathcal{M}dz} \left.\frac{dE_{\rm gw}}{d\ln f_r} \right|_{f_r=(1+z)f}\,,
\end{equation}
where $f_r$ and $f$ are the emitted and observed GW frequencies, and $d^2 n/d\mathcal{M}dz$ is the number density of sources per redshift and chirp mass bin.
The chirp mass is defined as $\mathcal{M}\equiv M_{\rm BH}[q/(1+q)^2]^{3/5}$ where $M_{\rm BH}\equiv M_1+M_2$ is the total mass of the binary and $q\equiv M_2/M_1$ is the mass ratio.
Assuming circular orbits, the energy spectrum is given by
\begin{equation}
    \frac{dE_{\rm gw}}{d\ln f_r} = \frac{1}{3G}(G\mathcal{M})^{5/3}(\pi f_r)^{2/3}\,,
\end{equation}
and the strain amplitude can be re-written as  \cite{Sato-Polito:2023gym,Liepold:2024woa}
\begin{eqnarray}
\label{eq:angle_avg_strain}
&&  \langle h^2 \rangle (f)=  \frac{1}{3c^2}\frac{1}{(\pi f)^{4/3}}  \\
        &&\times \int dM_{\rm BH}dqdz \frac{d^3n}{dM_{\rm BH}dqdz} \frac{q(GM_{\rm BH})^{5/3}}{(1+q)^2} \frac{1}{(1+z)^{1/3}}\,.\nonumber
\end{eqnarray}
Assuming that the number density distribution of SMBHB mergers factorizes as
\begin{equation}\label{eq:dn}
    \frac{d^3n}{dM_{\rm BH}dqdz} = p_z(z) p_q(q) \frac{dn}{dM_{\rm BH}}\,,
\end{equation}
where $p_z(z)$ and $p_q(q)$ are, respectively, normalized redshift and mass ratio distributions,
we then obtain Eq.~(5) of \cite{Liepold:2024woa} (note the factor of $4\pi$ difference between $\langle h^2 \rangle$ and $h_c^2$):
\begin{eqnarray}\label{eq:h2final}
    &&\langle h^2 \rangle (f) =  \frac{1}{3c^2(\pi f)^{4/3}} 
        \int dq \frac{q p_q(q)}{(1+q)^2} \int dz\frac{p_z(z)}{(1+z)^{1/3}} \nonumber\\
        &&\times \int dM_{\rm BH}(GM_{\rm BH})^{5/3}\frac{dn}{dM_{\rm BH}} \nonumber \\
    &=& 9.5\times10^{-32} \left(\frac{{\rm yr}^{-1}}{f}\right)^{4/3} 
        \int dq \frac{q p_q(q)}{(1+q)^2} \int dz\frac{p_z(z)}{(1+z)^{1/3}} \nonumber\\
        &&\times \int dM_{\rm BH} \left(\frac{M_{\rm BH}}{10^9M_\odot}\right)^{5/3} \frac{d\left(\frac{n}{10^{-4}{\rm Mpc}^{-3}}\right)}{dM_{\rm BH}}\,.
\end{eqnarray}

Extending this calculation to the fourth moment of the strain field $h$, we find an analogous expression to Eq.~\eqref{eq:h2c} for $\langle h^4 \rangle$:
\begin{eqnarray}\label{eq:h4}
    \langle h^4 \rangle(f) &=& \frac{1}{4\pi}\left(\frac{4G}{\pi c^2f^2}\right)^2 \int d\mathcal{M}\frac{dz}{(1+z)^2} \frac{d^2n}{d\mathcal{M}dz} \nonumber\\
        &\times&\left.\left(\frac{dE_{\rm gw}}{d\ln f_r}\right)^2\right|_{f_r=(1+z)f} \frac{d\ln f}{dV_c}\,.
\end{eqnarray}
The last term is a Jacobian factor relating the logarithmic frequency interval, $\ln f$, and comoving volume, $V_c$. It arises from the Dirac delta function constraint that the comoving spatial coordinates of two points must coincide for self-correlations. 
The Jacobian factor can be expressed as
\begin{eqnarray}
    \frac{d\ln f}{dV_c} &=& \frac{d\ln f}{d\ln f_r} \frac{d\ln f_r}{dt_r} \frac{dt_r}{dz} \frac{dz}{dV_c} \nonumber\\
    &=&\frac{d\ln f_r}{dt_r}\frac{1}{(1+z)4\pi c\chi^2}\,,
\end{eqnarray}
where $dt_r$ denotes a time interval measured in the source rest frame, $\chi$ is the comoving distance out to redshift $z$, and
\begin{equation}\label{eq:dfdt}
    \frac{d\ln f_r}{dt_r} = \frac{96}{5}\pi^{8/3} \left(\frac{G\mathcal{M}}{c^3} \right)^{5/3}f_r^{8/3}
\end{equation}
for circular orbits that decay solely due to GW emission \cite{Sesana:2008mz}.
Assuming the same factorized SMBHB merger model above, we obtain
\begin{eqnarray}\label{eq:h4final}
    &&\langle h^4 \rangle (f) = \left[\frac{1}{3c^2(\pi f)^{4/3}}\right]^2 \frac{96(\pi f)^{8/3}H_0^2}{5 c^8}     
        \int dq \frac{q^3 p_q(q)}{(1+q)^6}  \nonumber\\
        &&\times \int dz\frac{p_z(z)(1+z)}{(H_0\chi(z)/c)^2} 
        \int dM_{\rm BH}(GM_{\rm BH})^{15/3}\frac{dn}{dM_{\rm BH}} \nonumber \\
    &=& 3.0\times10^{-63} h_{70}^2 \int dq \frac{q^3 p_q(q)}{(1+q)^6} \int dz\frac{p_z(z)(1+z)}{(H_0\chi(z)/c)^2}  \nonumber\\ &&\times 
        \int dM_{\rm BH} \left(\frac{M_{\rm BH}}{10^9M_\odot}\right)^{5} \frac{d\left(\frac{n}{10^{-4}{\rm Mpc}^{-3}}\right)}{dM_{\rm BH}}\,,    
\end{eqnarray}
where we have inserted characteristic numbers, and $h_{70}$ denotes the Hubble parameter today in units of $70$ km/s/Mpc. Because the integral over mass is weighted as $\propto M_{\rm BH}^{5}$, we expect the shot-noise anisotropy to be sensitive to the high-mass tail of the SMBH mass function, as shown below.
Finally, the shot-noise power spectrum follows according to Eqs.~\eqref{eq:cl_nanograv}-\eqref{eq:clsn}:
\begin{widetext}
\begin{eqnarray}
    \frac{1}{4\pi}C_{\ell>0,h^2}^{\rm SN} &=& \frac{\langle h^4 \rangle}{4\pi\langle h^2 \rangle^2} =  \frac{24(\pi f)^{8/3}H_0^2}{5\pi c^8} 
        \frac{\int dq \frac{q^3 p_q(q)}{(1+q)^6}}{\left[\int dq \frac{q p_q(q)}{(1+q)^2}\right]^2}
        \frac{\int dz\frac{p_z(z)(1+z)}{(H_0\chi(z)/c)^2}}{\left[\int dz\frac{p_z(z)}{(1+z)^{1/3}}\right]^2}
        \frac{\int dM_{\rm BH}(GM_{\rm BH})^{5}\frac{dn}{dM_{\rm BH}}}{\left[\int dM_{\rm BH}(GM_{\rm BH})^{5/3}\frac{dn}{dM_{\rm BH}}\right]^2} \nonumber \\
    &=& 0.026 \, h_{70}^2 \left(\frac{f}{{\rm yr}^{-1}}\right)^{8/3} 
        \frac{\int dq \frac{q^3 p_q(q)}{(1+q)^6}}{\left[\int dq \frac{q p_q(q)}{(1+q)^2}\right]^2}
        \frac{\int dz\frac{p_z(z)(1+z)}{(H_0\chi(z)/c)^2}}{\left[\int dz\frac{p_z(z)}{(1+z)^{1/3}}\right]^2}
        \frac{\int dM_{\rm BH} \left(\frac{M_{\rm BH}}{10^9M_\odot}\right)^{5} \frac{d\left(\frac{n}{10^{-4}{\rm Mpc}^{-3}}\right)}{dM_{\rm BH}}}{\left[\int dM_{\rm BH} \left(\frac{M_{\rm BH}}{10^9M_\odot}\right)^{5/3} \frac{d\left(\frac{n}{10^{-4}{\rm Mpc}^{-3}}\right)}{dM_{\rm BH}}\right]^2}\,.
\end{eqnarray}
\end{widetext}

We highlight the steep frequency dependence
\begin{equation}\label{eq:ClSN-f}
        C_{\ell>0,h^2}^{\rm SN}\propto f^{8/3}\,,
\end{equation}
a direct consequence of Eq.~\eqref{eq:dfdt} under the assumption that SMBHBs lose orbital energy solely through gravitational wave emission. In this scenario, higher-frequency GWs are emitted at smaller orbital separations where the orbits decay more quickly. This means that fewer sources contribute to higher frequency bins, leading to larger shot-noise anisotropies.
    
More generally, non-GW processes such as interactions of SMBHBs with surrounding gas and/or stars can remove orbital energy from the binary system and lead to deviations from the above scaling. We use a power-law frequency-dependent ``residence time'' to characterize the amount of time that binaries spend emitting at different frequencies (e.g. \cite{Lamb:2024gbh}):
\begin{equation}
   dt_r/d\ln f_r \propto f^{-\beta} \,.
\end{equation}
The background mean strain amplitude scales as the product of the strain per source and the number of sources 
per $\ln f$, $\langle h^2 \rangle \propto h_s^2(f) \times dN/d\ln f$, where the single source orientation-averaged strain amplitude is $h_s^2(f) \propto ({dE_{\rm gw}}/{dt_r})/(f^2 d_L^2) \propto \mathcal{M}^{10/3} f^{4/3}/d_L^2$, and $d_L$ is the luminosity distance. 
The quantity $dE_{\rm gw}/dt_r$, and hence $h_s(f)$, depends only on the binary masses and orbital geometry and remains the same even in alternatives to purely GW-driven cases \cite{2011MNRAS.411.1467K}. The number of sources per $\ln f$ is proportional to the residence time $dN/d\ln f \propto dt_r/d\ln f_r$. Therefore, we have the following scaling relations:
\begin{equation}
        \langle h^2\rangle \propto h_s^2(f) \frac{dt_r}{d\ln f_r}\propto f^{4/3-\beta}\,.
\end{equation}
Similarly, the shot-noise anisotropy scales as 
\footnote{
The frequency dependence of the variance $\langle h^4\rangle$ here differs by a factor of $f$ from Table~1 of \cite{Lamb:2024gbh} because we adopt logarithmic frequency bins while they adopt linear frequency bins.}
\begin{equation}
    \langle h^4\rangle \propto h_s^4(f) \frac{dt_r}{d\ln f_r}\propto f^{8/3-\beta}\,,
\end{equation}
leading to
\begin{equation}\label{eq:SN-beta}
    C_{\ell>0,h^2}^{\rm SN} =      \frac{\langle h^4\rangle}{\langle h^2\rangle^2}\propto\frac{d\ln f_r}{dt_r}\propto f^{\beta}\,,
\end{equation}
where $\beta=8/3$ for GW-driven circular orbits. 
If the binaries spend less time at large separations (small frequencies) because of gas dynamical and/or stellar interactions, then fewer sources contribute to the low-frequency bands, and the shot-noise increases relative to the GW-driven scenarios, leading to a flatter spectrum with $\beta < 8/3$.
Therefore, measurements of the frequency dependence of the shot-noise anisotropies provide an additional avenue, beyond the frequency scaling of the mean spectrum, for determining SMBHB residence times \cite{Sato-Polito:2023spo}. 

\subsection{Anisotropies from Large-Scale Structure}
To compute the LSS anisotropy contributions to the GWB anisotropies, $C_{\ell>0,h^2}^{\rm LSS}$, we modify the spatially-averaged SMBHB merger rate, $d^3n/{dM_{\rm BH} \, dq \, dz}$, in Eq.~\eqref{eq:angle_avg_strain} to account for variations with angle on the sky and in redshift:
\begin{align}\label{eq:excess_merger_rate}
    \frac{d^3n(\boldsymbol{\hato}, z)}{dM_{\rm BH} \, dq \, dz} 
    &= \left[ 1 + b_{\rm BH}(M_{\rm BH}, q, z) \, 
        \delta_{\rm lin}(\boldsymbol{\hato}, z) \right] \nonumber \\
    &\quad \times \frac{d^3n}{dM_{\rm BH} \, dq \, dz} \,,
\end{align}
where $\delta_{\rm lin}(\boldsymbol{\hato},z)$ denotes the local linear matter density fluctuation
and $b_{\rm BH}(z,q,M_{\rm BH})$ is a bias factor relating the black hole merger rate in an overdense region to the local overdensity.
The bias factor depends on how the SMBHBs populate host dark matter halos and can be a function of $M_{\rm BH}, q$ and $z$.
We can then use Eq.~\eqref{eq:angle_avg_strain} to obtain the fractional fluctuation in the strain amplitude in direction $\boldsymbol{\hato}$:
\begin{eqnarray}
\label{eq:strain_flucs}
    \delta_{h^2}(\boldsymbol{\hato}) =  \frac{1}{\langle h^2 \rangle}\frac{1}{3c^2(\pi f)^{4/3}} 
        \int dz\frac{p_z(z)}{(1+z)^{1/3}} \int dq \frac{q p_q(q)}{(1+q)^2} \nonumber\\
    \times \int dM_{\rm BH}(GM_{\rm BH})^{5/3}\frac{dn}{dM_{\rm BH}} b_{\rm BH}(z,q,M_{\rm BH}) \delta_{\rm lin}(\boldsymbol{\hato}, z), \nonumber \\
\end{eqnarray}
where we have assumed the factorization of Eq.~\eqref{eq:dn} for simplicity. 
Defining a redshift-dependent effective bias factor
\begin{equation}\label{eq:beff}
    \langle b_{\rm BH}(z) \rangle \equiv 
    \frac{
    \int\! dM_{\rm BH} \, dq \, (G M_{\rm BH})^{5/3} \frac{dn}{dM_{\rm BH}} \frac{q\, p_q(q)}{(1+q)^2} b_{\rm BH}(z,q,M_{\rm BH})
    }{
    \int\! dM_{\rm BH} \, dq \, (G M_{\rm BH})^{5/3} \frac{dn}{dM_{\rm BH}} \frac{q\, p_q(q)}{(1+q)^2}
    }\,,
\end{equation}
we can simplify Eq.~\eqref{eq:strain_flucs} to
\begin{align}\label{eq:strain_fluc_final}
    \delta_{h^2}(\boldsymbol{\hato}) &= \frac{\int dz \frac{p_z(z)}{(1+z)^{1/3}} \langle b_{\rm BH}(z) \rangle \frac{D(z)}{D(0)} \delta_{\rm lin}(\boldsymbol{\hato},z=0)}{\int dz \frac{p_z(z)}{(1+z)^{1/3}}} \nonumber \\
    & \equiv \int dz \, \mathcal{P}_{h^2}(z) \langle b_{\rm BH}(z) \rangle \frac{D(z)}{D(0)} \delta_{\rm lin}(\boldsymbol{\hato},z=0)\,,
\end{align}
where $D(z)/D(0)$ is the linear growth factor normalized to unity today, and $\mathcal{P}_{h^2}(z)$ denotes the fraction of the strain signal produced between $z$ and $z + dz$. Eq.~\eqref{eq:strain_fluc_final} is analogous to the expression for the angular fluctuations in a photometric galaxy sample, where $\mathcal{P}_{h^2}(z)$ plays the role of the galaxy redshift distribution, and $\langle b_{\rm BH}(z) \rangle$ is analogous to the galaxy clustering bias, although here the relevant bias factor is strain amplitude-weighted, rather than number density-weighted. Note that the expression with $\mathcal{P}_{h^2}(z)$ applies to a more general merger rate model, including cases beyond the factorization assumed in Eq.~\eqref{eq:dn}.

The angular power spectrum of the fractional strain amplitude fluctuations follows directly from Eq.~\eqref{eq:strain_fluc_final} by analogy with standard LSS results (e.g., \cite{Tegmark:2001xb}). It can be obtained by expanding the linear density fluctuations in Fourier modes using the plane wave expansion
\begin{equation}
\label{eq:rayleigh_expansion}
    e^{i \boldsymbol{k \cdot x}} = 4 \pi \sum_{\ell, m} i^\ell j_\ell(k x) Y_{\ell m}(\boldsymbol{\hat{k}}) Y^*_{\ell m}(\boldsymbol{\hat{x}}),
\end{equation}
where the spherical Bessel functions $j_\ell$ act as projection kernels mapping 3D fluctuations onto angular modes on the sky, and the spherical harmonics describe the angular dependence. 
Using the orthogonality of the spherical harmonics, we obtain the standard result
\begin{equation}
\label{eq:cl_exact}
    C_{\ell,h^2}^{\rm LSS} = \frac{2}{\pi}\int \frac{dk}{k} k^3 P_{\rm lin}(k,z=0) [F_\ell^{h^2}(k)]^2,
\end{equation}
where $P_{\rm lin}(k,z=0)$ is the present-day linear matter power spectrum, and 
\begin{equation}
    F_\ell^{h^2}(k) = \frac{\int dz \frac{p_z(z)}{(1+z)^{1/3}} \langle b_{\rm BH}(z) \rangle \frac{D(z)}{D(0)} j_\ell[k\chi(z)]}{\int dz \frac{p_z(z)}{(1+z)^{1/3}}}
\end{equation}
is a kernel that encodes the line-of-sight projection of the SMBHB merger redshift distribution, weighted by the effective bias and the linear growth factor.

It is also useful to note the corresponding Limber approximation result \cite{Limber53,Kaiser92}:
\begin{eqnarray}
\label{eq:limber}
    &&C_{\ell,h^2}^{\rm LSS} = \int \frac{dz}{\chi^2(z)}\frac{H(z)}{c} \left[\frac{p_z(z)}{(1+z)^{1/3}} \langle b_{\rm BH}(z) \rangle \frac{D(z)}{D(0)}\right]^2 \nonumber\\
        &\times& P_{\rm lin}\left(k=\frac{\ell+1/2}{\chi(z)},z=0\right)/\left[\int dz \frac{p_z(z)}{(1+z)^{1/3}}\right]^2\,. \nonumber \\
\end{eqnarray}
The Limber limit applies when the Bessel kernels oscillate rapidly along the line-of-sight, so that short wavelength radial modes average out, and only transverse modes with $k_\perp \chi(z) \sim \ell + 1/2$ contribute significantly.
We will use Eqs.~\eqref{eq:cl_exact}-\eqref{eq:limber} to calculate the LSS-tracing anisotropies in empirical models for SMBHB mergers in the next section.

We note that the LSS piece of the GWB anisotropy spectrum here is independent of frequency, in contrast to the strong frequency dependence in the shot noise spectrum in Eq.~\eqref{eq:ClSN-f}. This is the case because large-scale overdensities simply rescale the merger rate (Eq.~\eqref{eq:excess_merger_rate}). The same population-averaged frequency dependence occurs in both the mean signal and the anisotropies, leaving the fractional strain fluctuations of Eq.~\eqref{eq:strain_flucs} independent of frequency.

\section{Results from empirical models}\label{sec:results}
\begin{figure}
    \centering
    \includegraphics[width=0.99\linewidth]{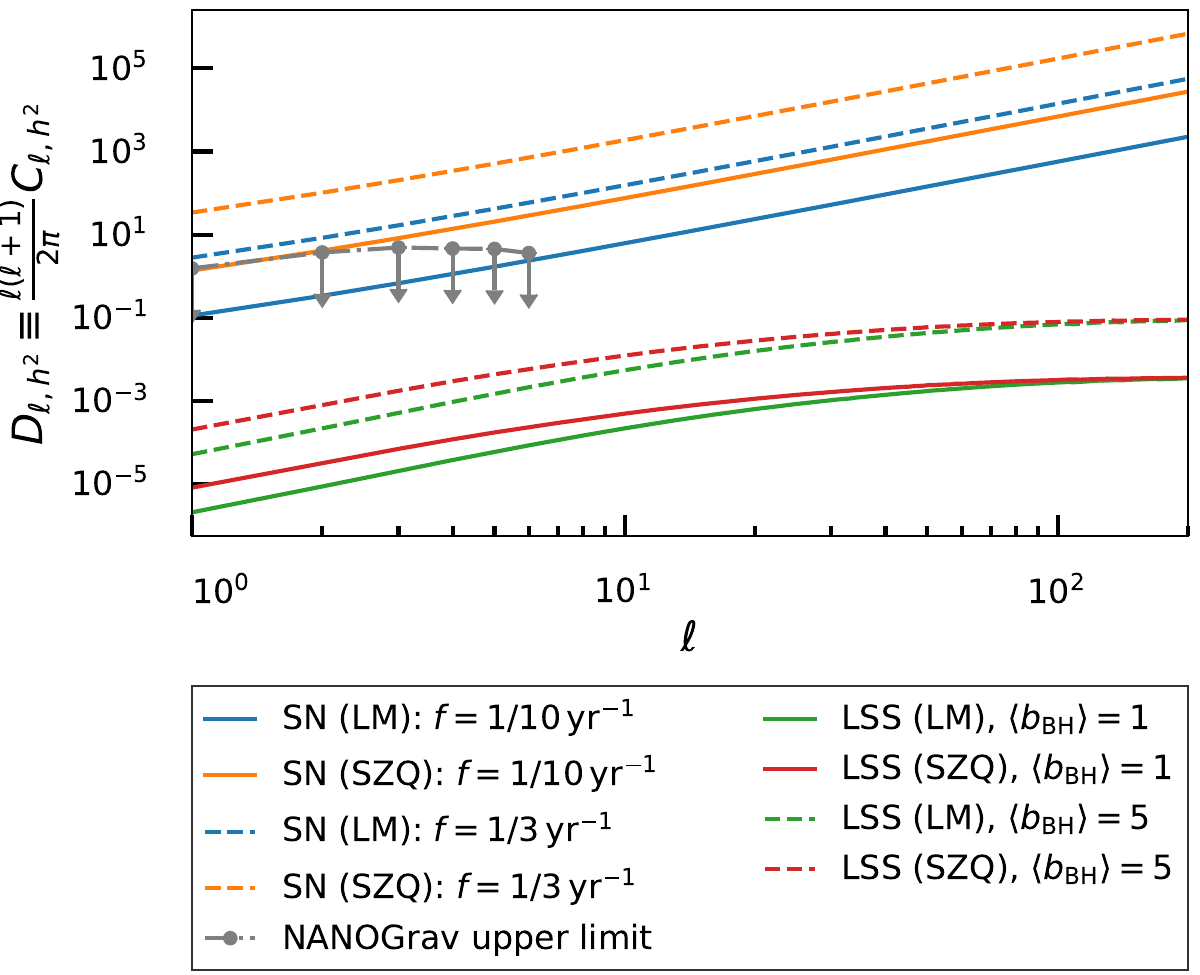}
    \caption{Model predictions for the GWB anisotropy power spectrum and current observational bounds. 
    The downward-pointing arrows show current 3-$\sigma$ upper bounds on the anisotropy signal from the NANOGrav collaboration based on a broad frequency-band analysis. The blue solid line gives the LM shot-noise model prediction at a frequency of $f=1/10 \, {\rm yr}^{-1}$, while the blue dashed line shows the LM shot-noise model at $f=1/3 \, {\rm yr}^{-1}$. The orange lines show the same for the SZQ model. 
    The quantity $D_{\ell, h^2}$ is plotted (see text), and so shot-noise scales as $\propto \ell^2$ here. The green and red solid lines show the LSS anisotropy power spectrum in the LM and SZQ models, respectively, with the solid lines adopting $\langle b_{\rm BH}(z) \rangle = 1$ and dashed lines taking $\langle b_{\rm BH}(z) \rangle=5$. Each of the shot-noise and LSS models assumes $z_{\rm min}=0.05$ and adopts a maximum mass of $M_{\rm BH, max}=10^{10.5} M_\odot$. 
    As we discuss below, the expected shot-noise signals are typically close to, or exceed, the NANOGrav bounds. Although a rigorous comparison between the models and the NANOGrav bounds is beyond the scope of this work, the results here suggest that shot-noise detections may be possible in the near term. The LSS signal is much smaller than the shot-noise one and the current observational bounds. 
    }
    \label{fig:Cl-signals}
\end{figure}

To make quantitative predictions for the expected shot-noise and LSS anisotropies, we consider two empirically calibrated models for the SMBHB merger rates:
\begin{itemize}
\item LM Model~\cite{Liepold:2024woa}: The SMBH mass function is obtained by convolution between the stellar mass function of the host galaxies, $dn/d M_*$, which quantifies the abundance of galaxies as a function of their stellar mass $M_*$, and a probability distribution function associated with the $M_{\rm BH}-M_*$ scaling relation:
\begin{equation}
    \frac{dn}{dM_{\rm BH}} = \int d M_*\frac{p(\log_{10} M_{\rm BH} | \log_{10} M_*)}{M_{\rm BH} \ln(10)} \frac{dn}{d M_*},
\end{equation}
where $p(\log_{10} M_{\rm BH} | \log_{10} M_*)$ is the conditional probability that a galaxy of mass $M_*$ hosts an SMBH of mass $M_{\rm BH}$. LM models this with a lognormal distribution whose median black hole mass is
\begin{equation}
    \log_{10} \left(\frac{M_{\rm BH}}{M_\odot}\right) =  \alpha + \beta  \log_{10} \left( \frac{M_*}{10^{11} M_\odot} \right)
\end{equation}
with $\alpha=8.46\pm 0.08$ and $\beta=1.05\pm 0.11$,
and whose scatter is $\epsilon_0 = 0.34$ dex in $\log_{10} M_{\rm BH}$ based on \cite{McConnell:2012hz}. The stellar mass function $dn/d M_*$ is given by Eq.~(2) of \cite{Liepold:2024woa}.
The redshift and mass-ratio distributions of SMBHB mergers are parameterized as
\begin{equation}\label{eq:lm_params}
    p_z(z) \propto z^\gamma e^{-(z/z_*)^2}\,,\quad p_q(q) \propto q^2\,
\end{equation} 
with $\gamma=1.0$, and $z_*=0.5$ based on the simulation results in~\cite{NANOGrav:2023hfp}.

\item SZQ Model~\cite{Sato-Polito:2023gym}: 
Instead of using galaxy stellar mass to infer $M_{\rm BH}$ as in the LM model, SZQ uses the velocity dispersion of the host galaxies
\begin{equation}
    \frac{dn}{dM_{\rm BH}} = \int d \sigma\frac{p(\log_{10} M_{\rm BH} | \log_{10} \sigma)}{M_{\rm BH} \ln(10)} \frac{dn}{d \sigma}\,.
\end{equation}
They model this with a lognormal distribution where the median black hole mass is
\begin{equation}
    \log_{10} \left(\frac{M_{\rm BH}}{M_\odot}\right) =  \alpha + \beta\,  \log_{10} \left( \frac{\sigma}{200\,{\rm km\,s^{-1}}} \right)
\end{equation}
with $\alpha=8.32\pm 0.05$ and $\beta=5.64\pm 0.32$,
and the scatter is $\epsilon_0 = 0.38$ dex in $\log_{10} M_{\rm BH}$. 
The velocity dispersion function is given by Eq.~(2) of \cite{Sato-Polito:2023gym}.
The merger distributions are parameterized as
\begin{equation}\label{eq:szq_params}
    p_z(z) \propto z^\gamma e^{-z/z_*}\,,\quad
    p_q(q) \propto q^{-1}\,
\end{equation} 
with $\gamma=0.5$, and $z_*=0.3$ chosen to approximately match the simulation results in~\cite{Kelley:2016gse}. 
\end{itemize}

The two models differ mainly in how the local SMBH mass function is inferred.
Additionally, the two models assume different redshift and mass-ratio distributions of SMBHB mergers, where the forms used in \cite{Liepold:2024woa} are calibrated to SMBHB population synthesis results from the NANOGrav collaboration.
These different choices reflect uncertainties in the underlying merger prescriptions, especially in the delay time distributions between galaxy and SMBHB mergers, and the mapping between galaxy and SMBH mass ratios.

We will see that the shot-noise model predictions are sensitive to the extreme high-mass end of the mass function, especially in the case of SZQ.
This may be partly an artifact of assuming simple functional forms for the high-mass tail of the mass function and for the scatter between $M_{\rm BH}$ and galaxy properties. 
Consequently, we generally adopt a sharp truncation in the SMBH mass function at a mass of $M_{\rm BH, max}$, with a fiducial value of $M_{\rm BH, max} = 10^{10.5} M_\odot$, comparable to the most massive black holes detected dynamically in the local universe (e.g., \cite{Liepold25}).

\begin{figure*}
    \centering
    \includegraphics[width=0.49\linewidth]{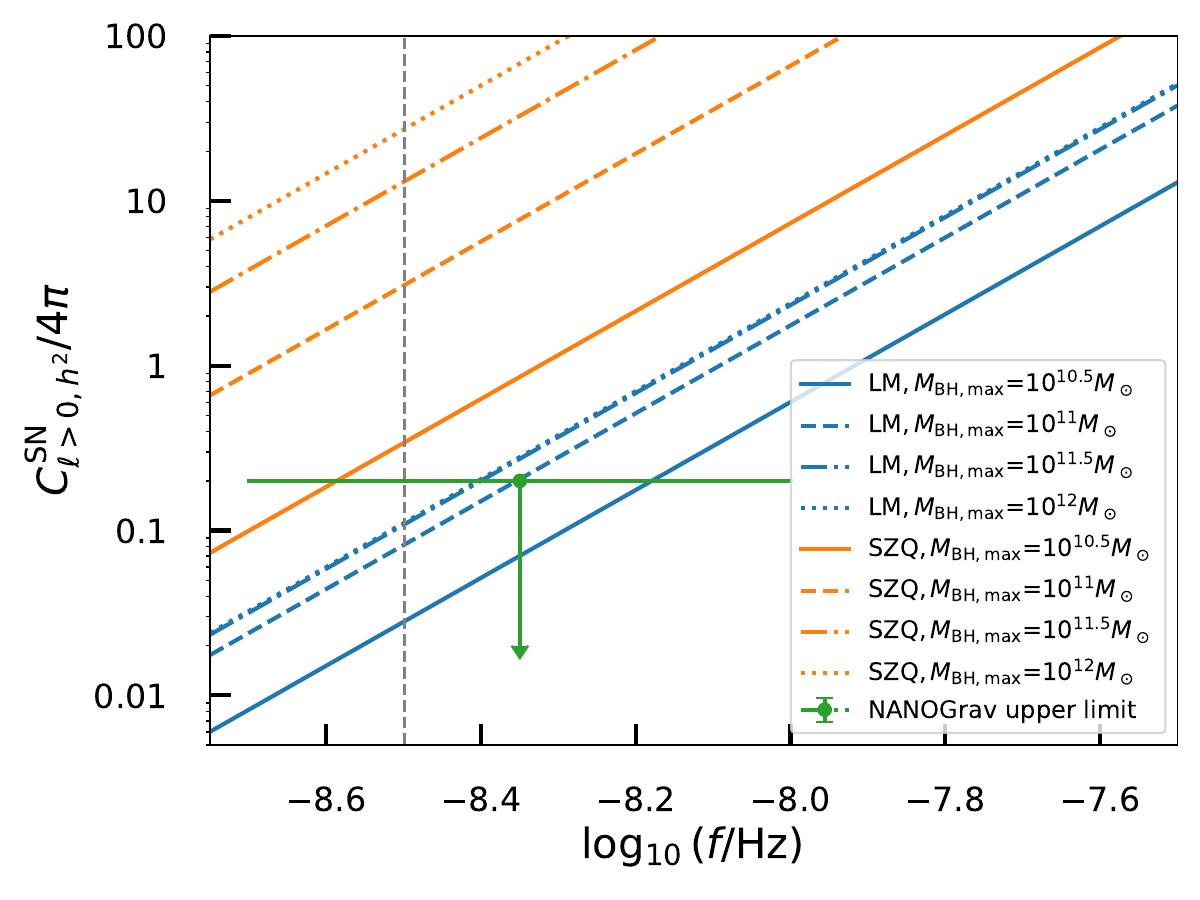}
    \includegraphics[width=0.49\linewidth]{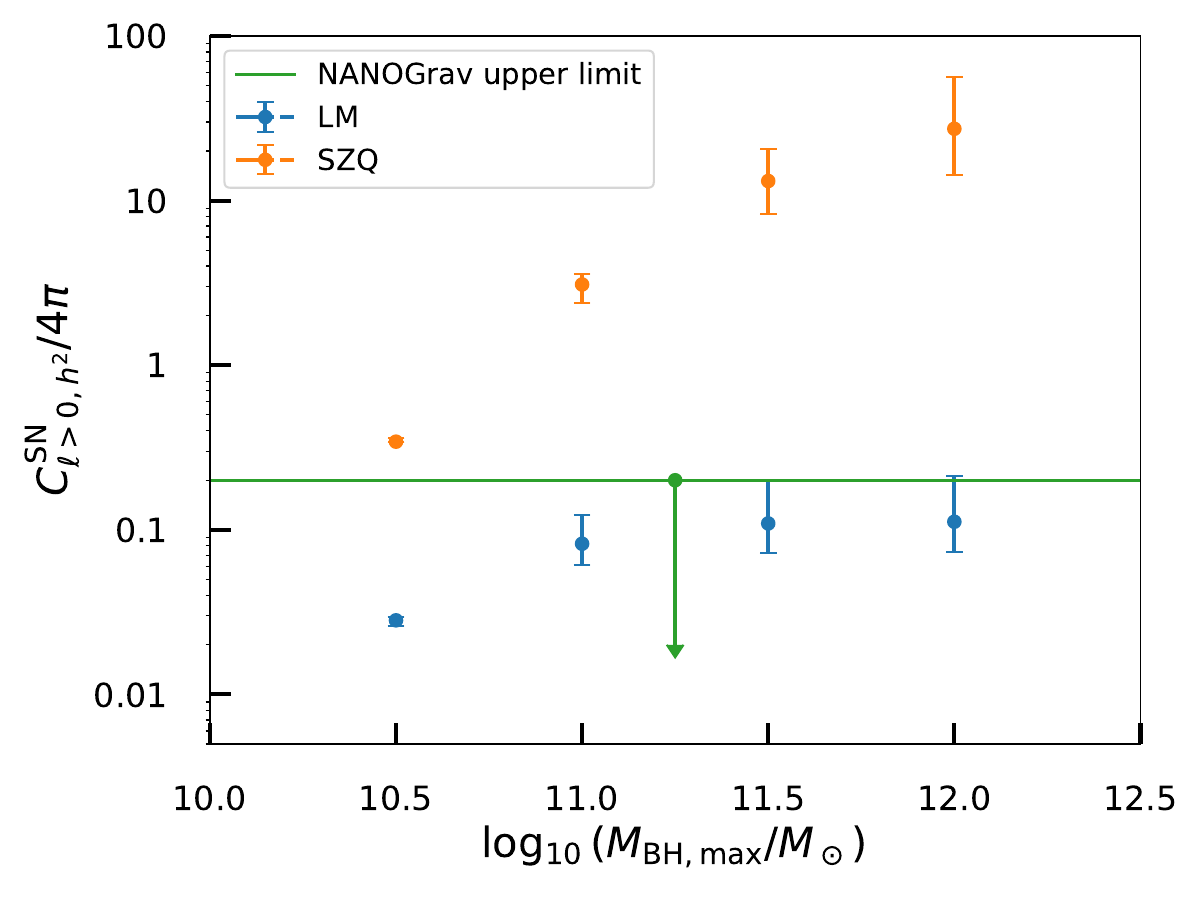}
    \caption{
    A comparison between the broad frequency-band NANOGrav constraint and the frequency dependence of the shot-noise models (left panel). The blue lines give the shot-noise power spectrum in the LM model, while the orange lines are for SZQ, each assuming a narrow logarithmic bin in frequency. 
    Each model takes $z_{\rm min} = 0.05$, while various values of $M_{\rm BH, max}$ are adopted, as indicated in the legend. 
    The green downward-pointing arrow shows the $3-\sigma$ upper bound from NANOGrav. The line indicates the approximate frequency range spanned in this broadband measurement, although we caution that the frequency weighting is non-uniform. 
    In each model, the shot-noise is a strongly increasing function of frequency owing to the decreasing residence time of SMBHB mergers towards high frequency. Future NANOGrav analyses adopting a frequency-dependent shot-noise template may help detect or constrain the shot-noise signal, providing further tests of our local SMBH census and of merger residence time models. 
    The results also reveal sensitivity to the high-mass tail of the SMBH mass function, especially in SZQ. 
    The right panel shows the $1-\sigma$ spread in shot-noise model predictions given the     
    SMBH mass function parameter uncertainties, for a few different values of $M_{\rm BH, max}$. Each case adopts $f=0.1 \, {\rm yr}^{-1}$ (denoted by the vertical dashed line in the left panel).  
    Note that, especially when $C_{\ell>0,h^2}^{\rm SN}/4\pi \gtrsim 1$, we expect a large sample variance around the mean shot-noise signal. 
    }
    \label{fig:SN-signals-f}
\end{figure*}

Fig.~\ref{fig:Cl-signals} shows predictions for both the shot-noise and LSS angular power spectra in each model, compared with current NANOGrav bounds on the anisotropy signals. Here we plot $D_{\ell,h^2} \equiv \ell (\ell + 1) C_{\ell,h^2}/(2 \pi)$,
which characterizes the variance of the fractional fluctuations in the strain amplitude at an angular scale of $\theta \sim 180^\circ/\ell$.
Each model assumes $z_{\rm min}=0.05$ and $M_{\rm BH, max}=10^{10.5} M_\odot$, while the LSS calculations consider constant bias factors, with $\langle b_{\rm BH}(z)\rangle = 1$ and $\langle b_{\rm BH}(z)\rangle = 5$ cases given for contrast, as discussed below and in the Appendix.  
The anisotropies due to shot noise are a strong function of frequency (Eq.~\eqref{eq:ClSN-f}), and so these results are shown at two example frequencies with $f=1/10\,{\rm yr}^{-1}$ and $1/3\,{\rm yr}^{-1}$, whereas the anisotropies due to LSS are independent of frequency, as noted above. In all cases considered, the expected shot-noise signals are at least two orders of magnitude larger in $C_{\ell,h^2}$ than the LSS-tracing anisotropies. The large shot-noise arises because the strain amplitudes in these models are dominated by rare bright sources, especially at high frequencies. The LSS signal is small because the strain signal traces a relatively large volume, and the universe is close to homogeneous and isotropic when averaged over such large scales. Note that the variance $D_{\ell,h^2}$ can exceed unity on small-scales. This implies that the strain field is skew-positive in this regime, since it is a positive definite quantity.

\subsection{Discussion of shot-noise signals}
\subsubsection{Sensitivity to rare massive sources}
To illustrate the sensitive dependence of the shot noise anisotropy power spectrum on $M_{\rm BH, max}$, we plot $C_{\ell>0,h^2}^{\rm SN}$ predicted by the LM and SZQ models for a range of $M_{\rm BH, max}$ in Fig.~\ref{fig:SN-signals-f}.
The lowest value shown, $M_{\rm BH, max}=10^{10.5} M_\odot$, corresponds to the most massive SMBHs that have been detected dynamically in the local universe (e.g., \cite{Liepold25, McConnell:2011mu}). Predictions for higher mass cutoffs are also shown for comparison, but we note that SMBHs with $M_{\rm BH} \gtrsim 10^{11} M_\odot$ would be very rare or possibly absent in a $\Lambda$CDM universe. The masses of local SMBHs are on average $\sim 0.2\%$ of the stellar mass of the bulge component of the host galaxy (e.g., \cite{McConnell:2012hz}). SMBHs with $M_{\rm BH}\gtrsim 10^{11} M_\odot$ would 
therefore either reside in host galaxies with extreme bulge stellar masses of $\gtrsim 5\times 10^{13} M_\odot$, or be significant outliers of the local black hole and galaxy bulge mass scaling relation. Either case would require extreme formation or accretion processes to reach such high masses.

\begin{figure*}
    \centering
    \includegraphics[width=0.49\linewidth]{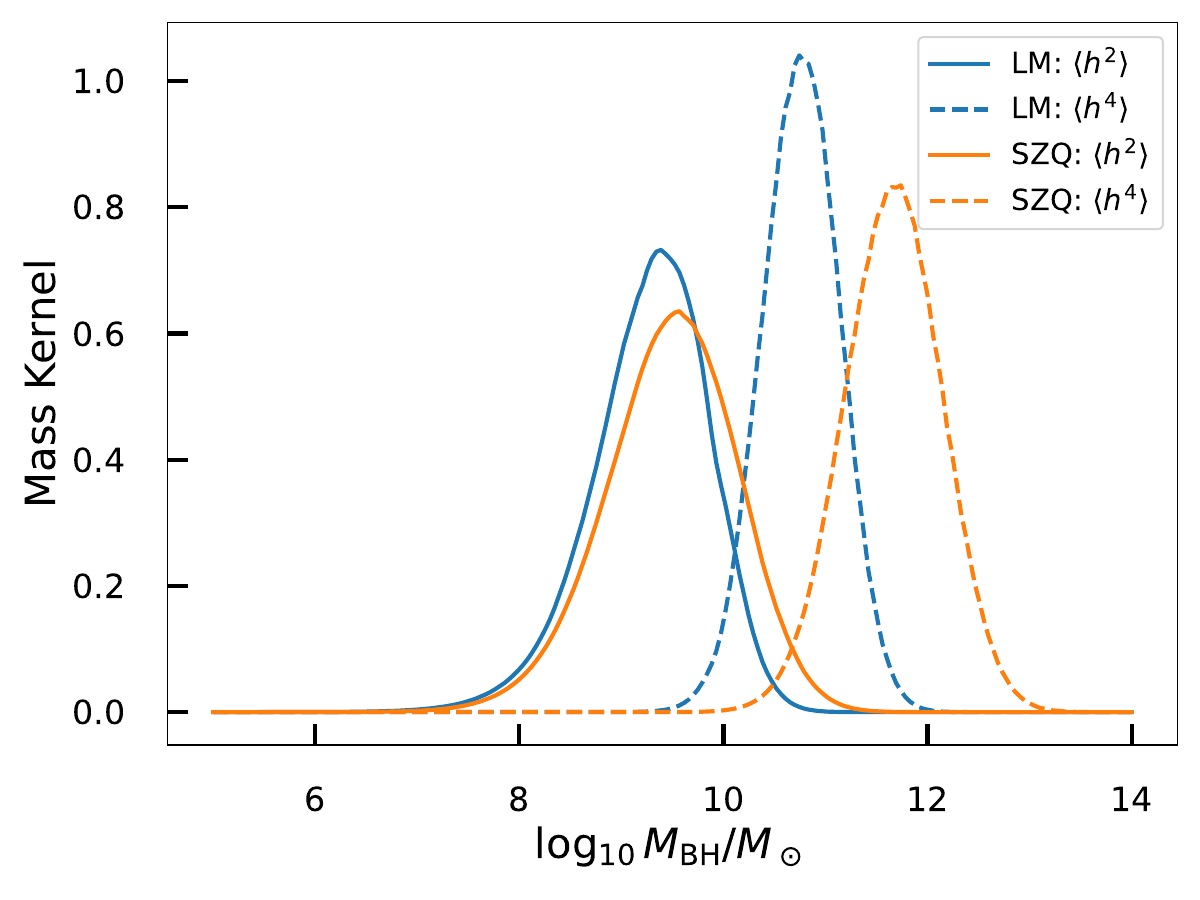}
    \includegraphics[width=0.49\linewidth]{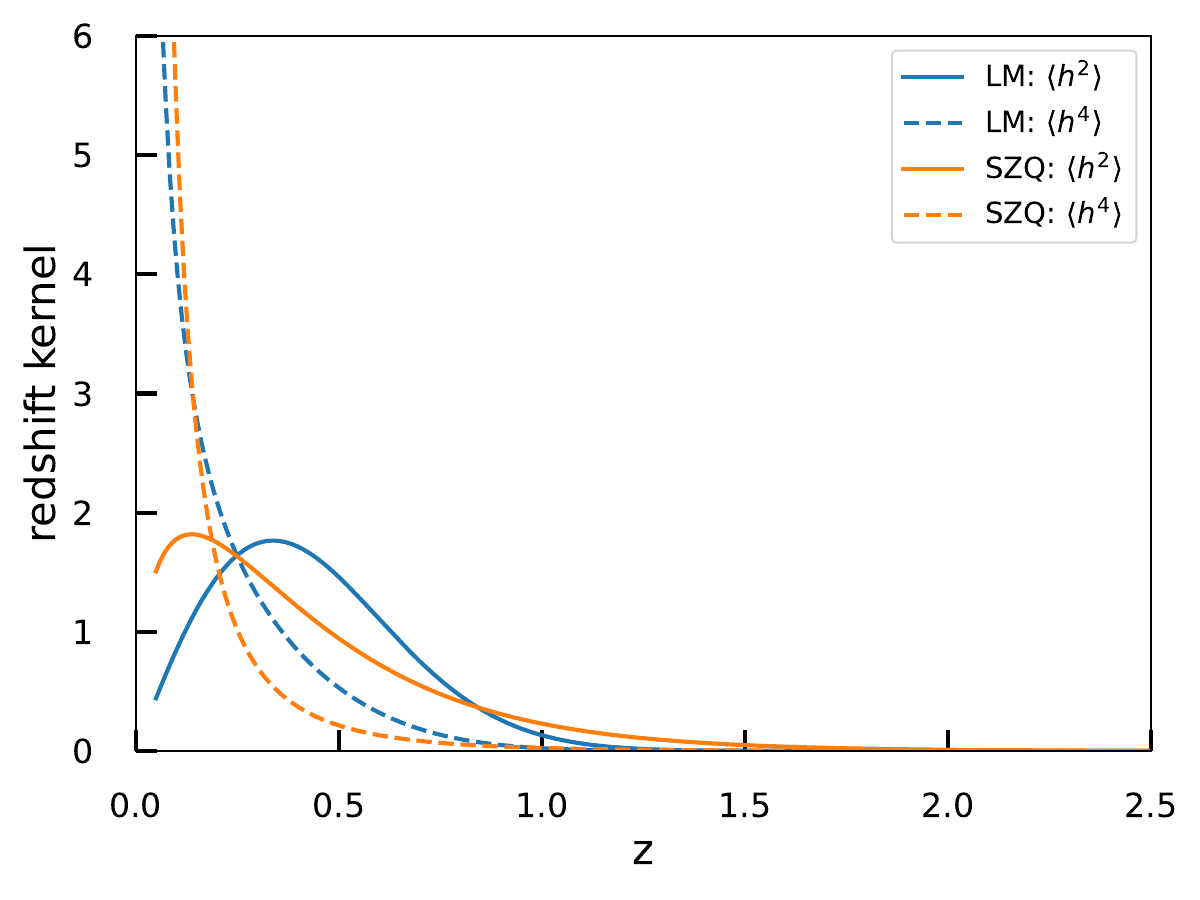}
    \caption{Contribution to the mean GWB ($\langle h^2 \rangle$, solid lines) and shot-noise induced anisotropy ($\langle h^4 \rangle$, dashed lines) per logarithmic mass bin (left panel), without adopting an $M_{\rm BH,max}$, and per redshift bin (right panel). All curves are normalized such that they integrate to unity.
    The blue lines are for the LM model, and the orange lines are for SZQ. The shot-noise anisotropy is weighted towards higher mass and lower redshift than the mean GWB signal, and so is sensitive to the extreme high-mass end of the SMBH function and to the distribution of nearby sources. The $\langle h^4 \rangle$ curve peaks at $z=z_{\rm min} =0.05$, reaching a redshift kernel value of 7.9 for LM and 16.8 for SZQ, exceeding the y-axis range shown. 
    }
    \label{fig:density}
\end{figure*}

Fig.~\ref{fig:SN-signals-f} shows significant differences between the two models: LM predicts an order of magnitude lower amplitude than SZQ for $M_{\rm BH, max}=10^{10.5} M_\odot$, and the difference between the two models increases to a factor of $\sim 300$ at $M_{\rm BH, max}=10^{12} M_\odot$.
The right panel of Fig.~\ref{fig:SN-signals-f} shows that the $C_{\ell>0,h^2}^{\rm SN}$ amplitude predicted by the LM model converges at $M_{\rm BH, max}\sim  10^{11} M_\odot$ while the SZQ model continues to rise with increasing $M_{\rm BH, max}$. The error bars here represent the uncertainties in the input SMBH mass functions, which become fairly large at high $M_{\rm BH, max}$. 

To quantify further the origin of these differences, we examine separately the mean GWB, $\langle h^2 \rangle$, and the shot-noise induced anisotropy, $\langle h^4 \rangle$, both of which enter the calculation of the power spectrum $C_{\ell>0,h^2}^{\rm SN}$. 
Fig.~\ref{fig:density} (left panel) compares the contributions per logarithmic mass bin to each of these two terms in the two models. 
Because $\langle h^4 \rangle$ involves an $M_{\rm BH}^5$-weighted average over the SMBH mass function (Eq.~\eqref{eq:h4final}), which is a much steeper function in black hole mass than the $M_{\rm BH}^{5/3}$ factor in $\langle h^2 \rangle$ (Eq.~\eqref{eq:h2final}), the mass contributions in the two models differ much more for $\langle h^4 \rangle$ (dashed curves) than for $\langle h^2\rangle$ (solid curves).
Specifically, the peak contributions to $\langle h^4 \rangle$ come from masses around $M_{\rm BH} \sim 10^{10.5} M_\odot$ in LM and $M_{\rm BH} \sim 10^{11.5} M_\odot$ in SZQ. This difference arises because the SZQ mass function has a more prominent tail towards high masses. 
The model predictions are also sensitive to the amount of scatter assumed in the $M_{\rm BH}-M_\star$ or $M_{\rm BH}-\sigma$ correlations, $\epsilon_0$, because this quantity affects the high mass tail of the SMBH mass function.
These results suggest that GWB anisotropy measurements could help inform the high-mass end of the SMBH mass function.

\begin{figure*}
    \centering
    \includegraphics[width=0.49\linewidth]{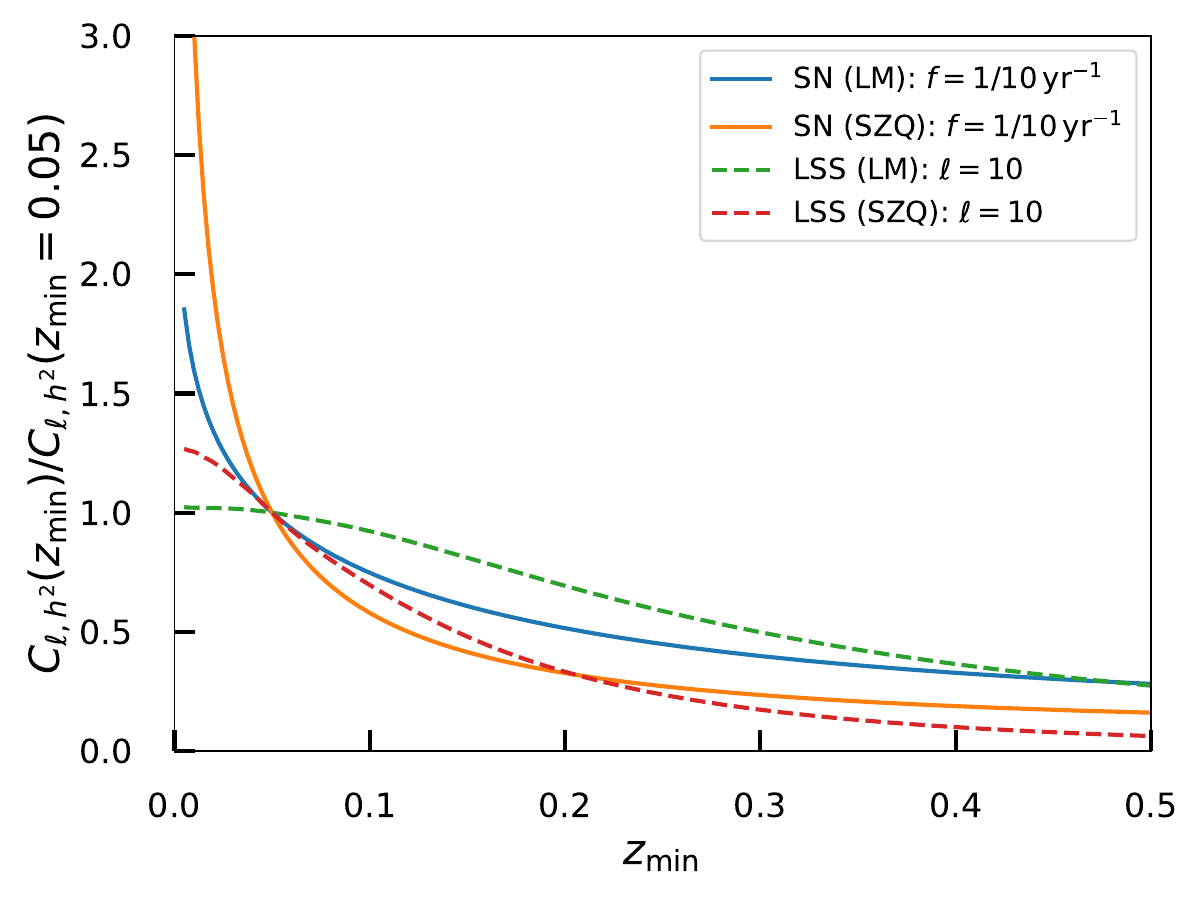}
    \includegraphics[width=0.49\linewidth]{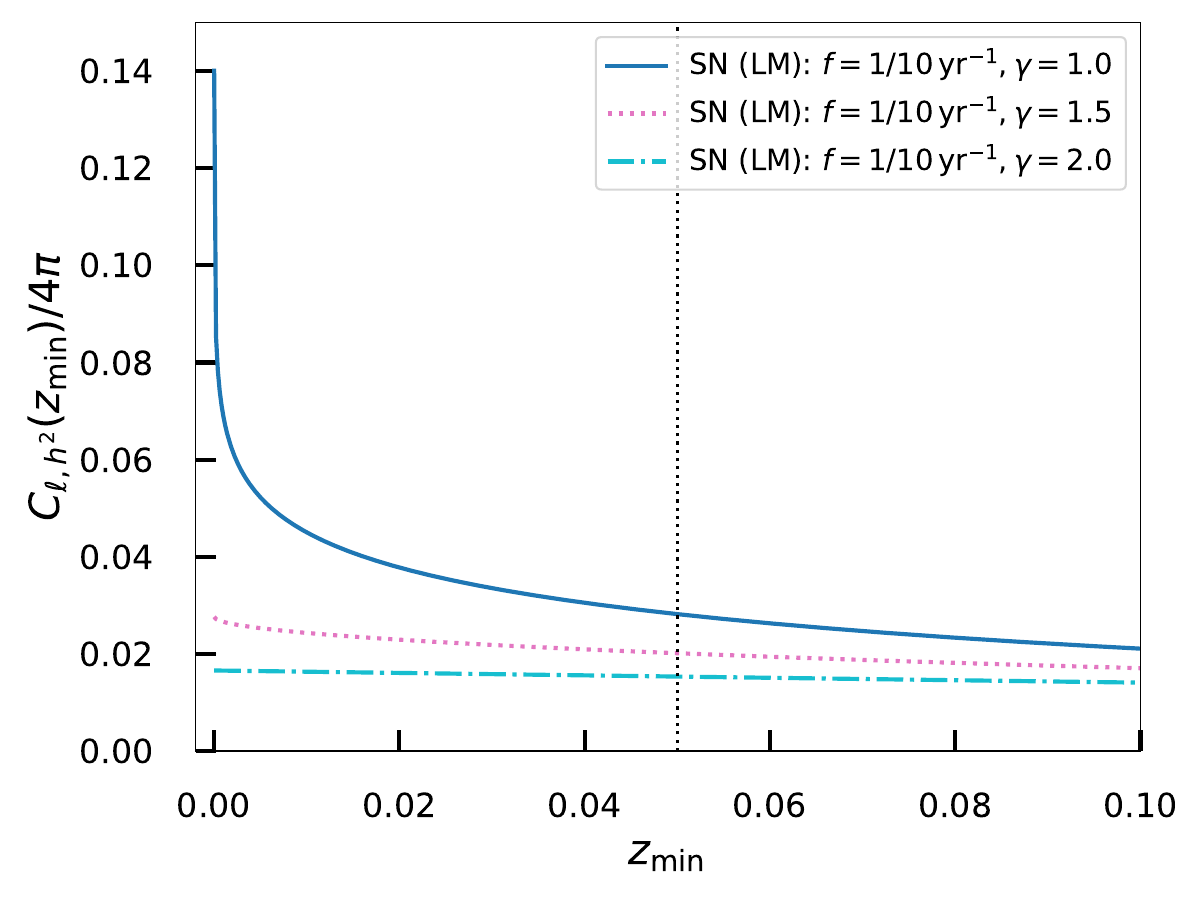}
    \caption{{\it Left panel:} the minimum redshift-dependence of anisotropies from both shot-noise and LSS, normalized to their values at $z_{\rm min}=0.05$. The blue and orange lines show, respectively, the LM and SZQ shot-noise power spectrum models at $f=1/10 \, {\rm yr}^{-1}$, while the green and red curves show the corresponding LM and SZQ LSS predictions at $\ell=10$. 
    The shot-noise power spectra formally diverge as $z \rightarrow 0$, and so there is some sensitivity to the precise choice of $z_{\rm min}$ assumed. 
    {\it Right panel:} the sensitivity of the shot-noise anisotropies to the redshift distribution power-law index, $\gamma$. For the LM model (here with $M_{\rm BH,max}=10^{10.5} M_\odot$) and $\gamma > 1$, the results converge towards $z_{\rm min} \rightarrow 0$, while the shot-noise amplitude drops with increasing $\gamma$.
    }
    \label{fig:zmin}
\end{figure*}

The green arrows in Fig.~\ref{fig:SN-signals-f} indicate the current NANOGrav $3\sigma$ upper bound on the GWB anisotropy from combining measurements at different frequencies across the approximate range shown \cite{NANOGrav:2023tcn}. Interestingly, the models generally exceed the NANOGrav bounds, especially in the case of SZQ and at high frequencies.
A proper comparison with NANOGrav data would require accounting for the non-trivial frequency-weighting in the measurement, which is beyond the scope of this work. 
One factor not considered here is the significant {\em sample variance} in the shot noise in addition to the contributions from the uncertainties in the SMBH mass function and merger rate distributions discussed earlier.
That is, the shot noise in different frequency bins should be regarded as random draws from an underlying probability distribution of shot-noise amplitudes. In scenarios where rare bright sources dominate the strain, the realization-to-realization scatter can be substantial: our universe may not be representative of the ensemble average, and different frequency bins effectively probe different random draws from the source population.
We explore this further in an upcoming follow-up paper~\cite{Lin:2026qvc}.

This situation closely parallels the case of the mean strain amplitude spectrum, $h^2_c(f)$, where rare bright sources generate a broad probability distribution of strains, and the median lies below the mean expected strain \cite{Sato-Polito:2024lew,Lamb:2024gbh}. We likewise expect substantial sample variance across different $\ell$-modes, as well as mode-to-mode correlations owing to the non-Gaussianity of the shot-noise signal. 
Additionally, the coherent summation of a limited number of binaries within each frequency bin can further enhance the variance relative to Gaussian expectations~\cite{Allen:2022dzg,Konstandin:2024fyo}.
In deriving constraints on the anisotropy spectrum from PTA measurements, it will therefore be important to model the full probability distribution of shot-noise amplitudes, and the covariance between $\ell$-modes, while also moving beyond the broad frequency-band upper bound in Fig.~\ref{fig:SN-signals-f}.

We caution that in the more extreme cases considered here, with $M_{\rm BH, max} \sim 10^{12} M_\odot$, even the GWs emitted near the innermost stable circular orbit would fall beneath the lower-frequency edge of the PTA band, while the earlier stages of the inspiral produce still lower frequency GWs. 
Therefore, even if such high mass SMBHBs exist, they are undetectable in PTAs. The high $M_{\rm BH, max} \sim 10^{12} M_\odot$ cases should hence be taken with a grain of salt, but they serve to illustrate the dependence on the high mass tail of the SMBH mass function.

\begin{figure}
    \centering
    \includegraphics[width=0.99\linewidth]{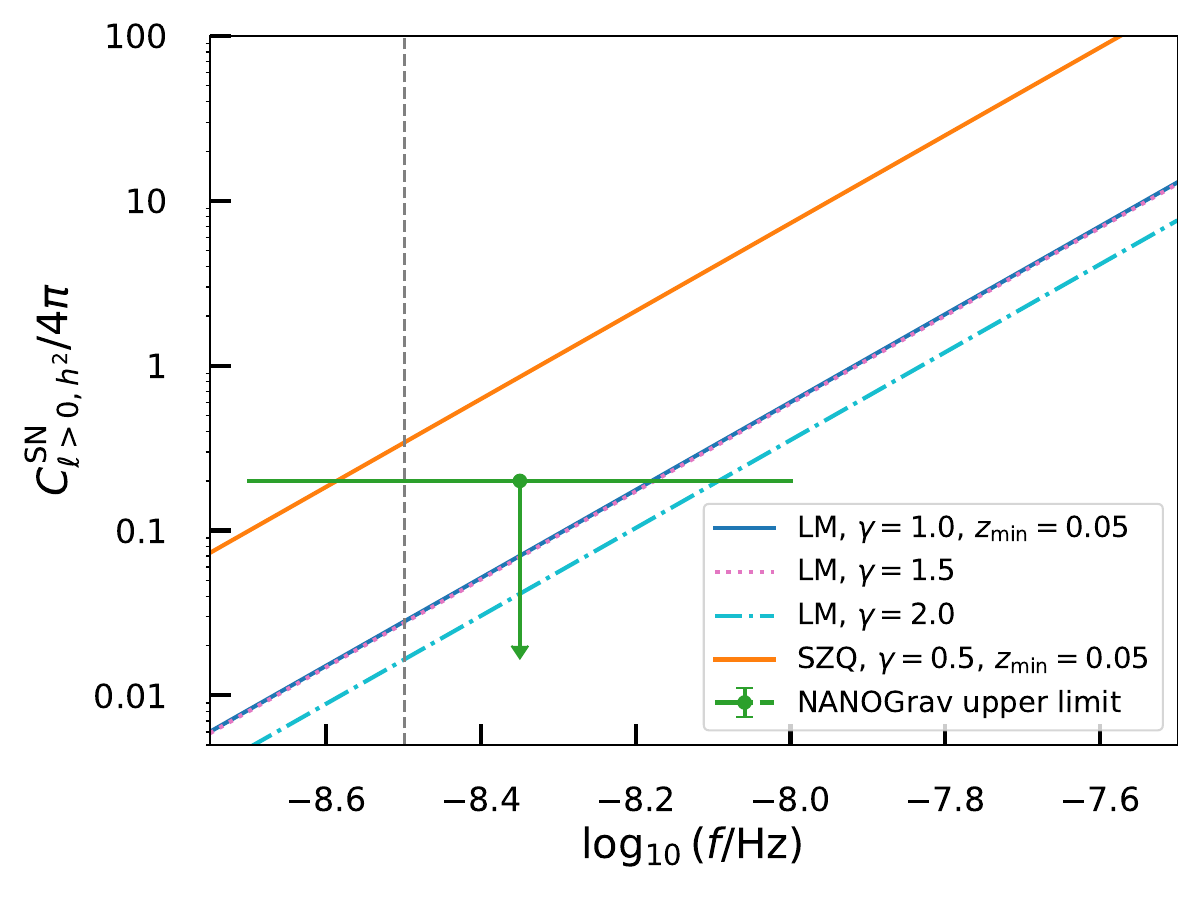}
    \caption{
    Similar to Fig.~\ref{fig:SN-signals-f} but with varying redshift distribution power-law index, $\gamma$, for the SMBH merger rates in the LM model.  The original LM and SZQ models are also shown for comparison (blue and orange lines), where the redshift integral
    in Eq.~\eqref{eq:h4zmin} is cut off at $z_{\rm min}=0.05$.  
    The modified LM models with $\gamma=1.5$ and $2.0$ are integrated to redshift 0. 
    All models assume $M_{\mathrm{BH,max}}=10^{10.5} M_\odot$. 
    }
    \label{fig:SN-gamma}
\end{figure}

\subsubsection{Sensitivity to low-redshift sources}

The redshift factor in the mean GWB amplitude $\langle h^2 \rangle$ in Eq.~\eqref{eq:h2final} depends very weakly on the exact form of the redshift distribution, $p_z(z)$, assumed for the SMBHB merger rates. For the LM and SZQ models, e.g., $\int dz\, p_z(z)/(1+z)^{1/3} = 0.890$ and $0.894$, respectively.

For the GWB anisotropy due to shot noise, however, we find the redshift factor in the expression for $\langle h^4 \rangle$ in Eq.~\eqref{eq:h4final}
\begin{equation}\label{eq:h4zmin}
    \langle h^4\rangle \propto \int dz \frac{p_z(z)(1+z)}{\chi^2(z)}
\end{equation}
to depend sensitively on the low-$z$ behavior of $p_z(z)$. This is because at $z\ll 1$, the comoving distance is $\chi(z) \approx z$, and $p_z(z) \propto z^\gamma$ in the LM and SZQ models.  
In fact, for $\gamma=1.0$ and $0.5$ assumed in the the two models, this integral diverges as $z \to 0$, as illustrated by the dashed curves in the right panel of Fig.~\ref{fig:density}.
For these two models, we impose a low-$z$ cutoff, $z_{\rm min}$, when computing the integral in Eq.~\eqref{eq:h4zmin}.
The left panel of Fig.~\ref{fig:zmin} illustrates the rapid increase in the resulting shot-noise anisotropy amplitude as $z_{\rm min}$ is decreased towards 0.

Since the redshift dependence of SMBH merger rates is poorly known observationally, we consider here $p_z(z)\propto z^\gamma$ with $\gamma >1$ that would yield convergent results without the need of applying an arbitrary $z_{\rm min}$.
The right panel of Fig.~\ref{fig:zmin} shows that the shot-noise anisotropies do not blow up at low redshifts when $\gamma$ is moved away from $1.0$.
The corresponding frequency-dependent shot-noise anisotropy spectrum is shown in Fig.~\ref{fig:SN-gamma}.  
The anisotropy amplitude is lowered by a factor of $\sim 2$ as $\gamma$ is increased to $2.0$.

The frequency-to-frequency scatter in the mean GWB measurements can be used to bound scenarios where the background becomes dominated by rare sources \cite{Sato-Polito:2024lew, Lamb:2024gbh}.
Here, our focus is the GWB anisotropies.
The shot-noise anisotropies provide another signature, along with the frequency-to-frequency scatter in $h^2_c(f)$, of rare bright source models.

\subsection{Discussion of LSS signals}

\begin{figure}
    \centering
    \includegraphics[width=0.99\linewidth]{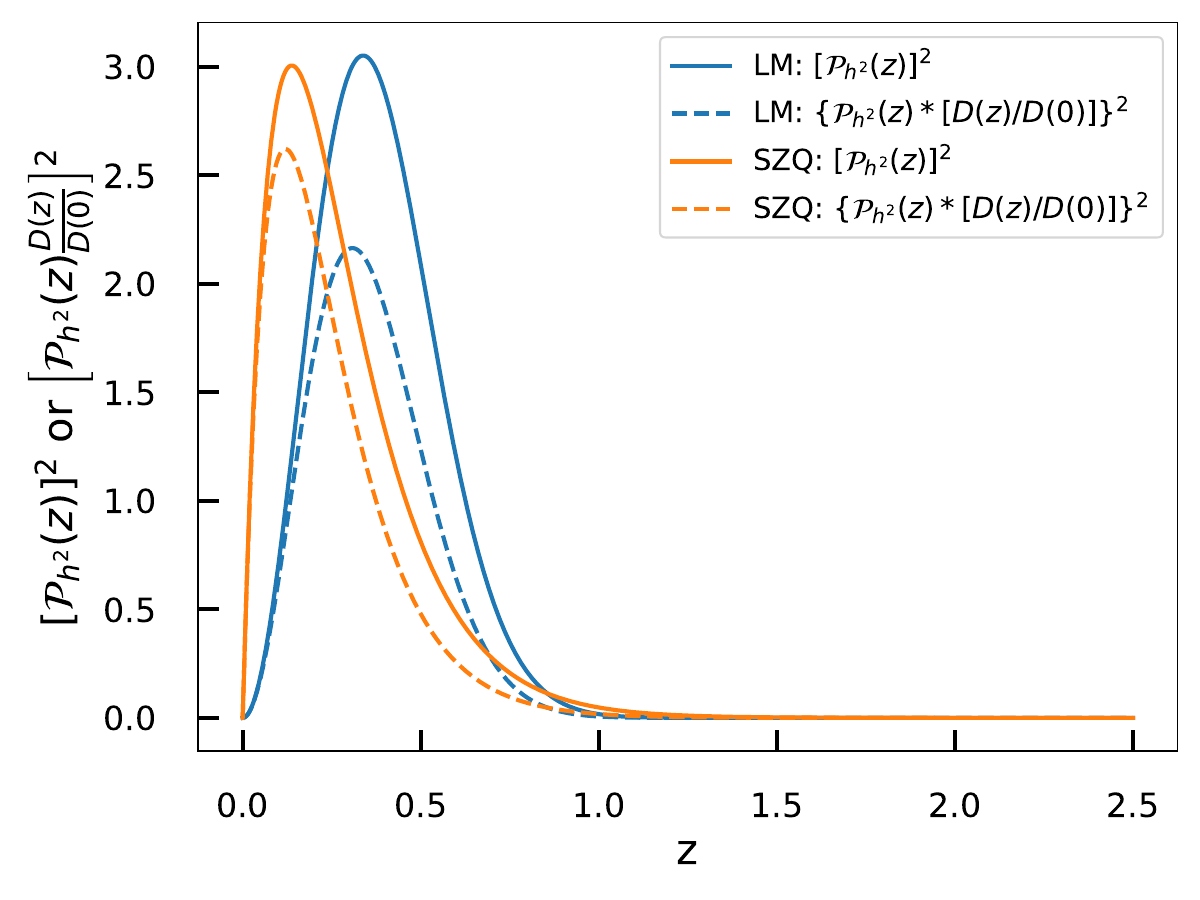}
    \caption{
    Contribution to the GWB anisotropy due to large scale structure (see Eqs.~\eqref{eq:cl_exact} and \eqref{eq:limber}) per redshift interval, where $\mathcal{P}_{h^2}(z)$ is defined in Eq.~\eqref{eq:strain_fluc_final} and $D(z)/D(0)$ is the cosmological linear growth factor.
    The larger angular power spectrum in the SZQ model (see Fig.~\ref{fig:Cl-signals}) arises because the mean signal peaks at a smaller redshift in SZQ than in LM. 
    }
    \label{fig:Wz}
\end{figure}

Under the redshift-independent bias factor assumption and the Limber approximation, the redshift integral in Eq.~\eqref{eq:limber} is determined by $\mathcal{P}_{h^2}(z) \propto p_z(z)/(1+z)^{1/3}$ and the linear growth factor, $D(z)/D(0)$. The squares of these quantities are given for the SZQ and LM models in Fig.~\ref{fig:Wz}. In each case, the redshift distribution falls off sharply above $z \gtrsim 1-2$, since SMBHB mergers that are sufficiently massive to produce reasonable strain signals in the PTA band are expected to be exceedingly rare at such high redshifts \cite{NANOGrav:2023hfp}. 
The SZQ model distribution peaks at lower redshift than in the LM model, while it also has a slightly more extended tail toward high redshift. As illustrated in the figure, the linear growth factor plays a minor role. This is the case because dark energy slows the growth of structure towards late times and $D(z)/D(0)$ is a weak function of redshift around the peak in $\mathcal{P}_{h^2}(z) \propto p_z(z)/(1+z)^{1/3}$. 

We note that the Limber approximation is accurate to better than 10\% across the full $\ell$-range and set of models shown in Fig.~\ref{fig:Cl-signals}.
This is a consequence of the relatively broad redshift distributions for the models in Fig.~\ref{fig:Wz}. This also reflects, in part, the smoothness of the linear matter power spectrum, which is also a slowly varying function of $k$ around its peak. 

The fact that the SZQ kernel peaks at lower redshifts than in the LM model leads to a larger angular power spectrum at low $\ell$ in the SZQ case, as apparent in Fig.~\ref{fig:Cl-signals}. Quantitatively, at $\ell=1$ the angular power spectrum is larger in SZQ than in LM by a factor of $4$ for $z_{\rm  min}=0.05$. 
Note that the LSS anisotropies also depend somewhat on $z_{\rm min}$, as illustrated in the left panel of Fig.~\ref{fig:zmin}.

The LSS signal also depends on the clustering bias of the SMBHB mergers. In Fig.~\ref{fig:Cl-signals} we contrast two cases in each model, with redshift-independent effective bias factors of $\langle b_{\rm BH} \rangle =1$ and $5$ (see Eq.~\eqref{eq:beff}). 
In the Appendix, we discuss a more detailed model for the clustering bias, which accounts for redshift evolution, and also compare with plausible SMBH-host galaxy clustering measurements. There we argue that the simpler, redshift-independent $\langle b_{\rm BH} \rangle = 1$ and $5$ models bracket the likely range of possibilities. 
As the angular power spectrum scales with $\langle b_{\rm BH} \rangle^2$ (Eq.~\eqref{eq:cl_exact}), this range translates into a factor of $25$ spread in the resulting angular power spectra. Even in the most optimistic case of $\langle b_{\rm BH} \rangle=5$ and the SZQ model, the LSS angular power spectrum is more than two orders of magnitude smaller in amplitude than the shot-noise at $f = 0.1\,{\rm yr}^{-1}$. The $\ell$-dependence of the LSS angular power spectrum is relatively weak at the most observationally accessible angular scales below $\ell \lesssim 10$. The peak in the angular power spectrum, $C_{\ell,h^2}^{\rm LSS}$, occurs near $\ell \sim 10$ in the models considered here.
This reflects the corresponding maximum in the matter power spectrum, $P_{\rm lin}(k)$, at $k \sim \ell/\chi \sim 0.01-0.02$ h/Mpc, for $\chi$ evaluated at redshifts close to the peak in the redshift distribution, near $z \sim 0.3-0.5$ (Fig.~\ref{fig:Wz}).   

Although the LSS signal is highly subdominant compared to the shot-noise fluctuations, we note that the signal may not be as small as one might naively expect. 
In the largest amplitude SZQ and $\langle b_{\rm BH} \rangle = 5$ scenario, $\sqrt{D_{\ell,h^2}^{\rm LSS}} = 0.015, 0.11, 0.30$ at $\ell =1$, $10$, and $100$, respectively. The relatively large fractional strain fluctuations, reaching an rms of 30\% at $\ell = 100 \sim 2^\circ$, are a consequence of the redshift distribution peaking around $z \sim 0.3$ in SZQ with considerable support at still lower redshifts. 
For contrast, in the case of LM and $\langle b_{\rm BH} \rangle =1$ the corresponding numbers are:  $\sqrt{D_{\ell,h^2}^{\rm LSS}} = 0.0014, 0.015, 0.053$ at $\ell =1$, $10$, and $100$, respectively, and so the model uncertainties remain large.
In any case, it will likely remain challenging to detect this signal in the foreseeable future.

\section{Conclusions and Discussions}\label{sec:conclusions}
The recent strong evidence for a GWB in PTA surveys motivates developing new methods to extract information about its origin.  
One key frontier is the search for spatial anisotropies in the GWB. 
Towards this end, we presented analytic calculations for GWB anisotropy signals, considering both shot-noise from the discrete merging SMBHB sources, as well as LSS-tracing terms. 

The shot-noise contribution to the angular power spectrum of the strain fluctuations is at least two orders of magnitude larger than the LSS signal. The amplitudes of the expected shot-noise signals are typically comparable to, or exceed, the current broad frequency-band NANOGrav constraints.
The shot-noise power spectrum increases sharply with frequency, scaling as $C_{\ell>0,h^2}^{\rm SN} \propto f^{8/3}$ across narrow logarithmic frequency bins, reflecting the decreasing residence time of merging black holes at high frequencies. 
Near-term PTA anisotropy analyses could therefore consider a two-parameter fit with a scale-independent shot-noise amplitude and a power-law index in frequency. This may facilitate detections or lead to tight bounds on our models. Departures from the $f^{8/3}$ scaling could indicate gas and/or stellar dynamical interactions, beyond simple GW-driven inspiral. 
Realization-to-realization scatter can also introduce deviations from the $f^{8/3}$ frequency scaling for the mean shot noise found here.
This is because the shot-noise signal is dominated by rare events at the high-mass end of the SMBH population, with peak contributions from masses around $M_{\rm BH} \sim 10^{10.5} M_\odot$ or beyond, depending on the precise mass function model. An important next step for deriving bounds on SMBHB models from upcoming shot-noise measurements will be to model the full probability distribution of the expected shot-noise amplitudes, accounting for the likely large sample variance, and the skewness of the distribution. A detailed study on this topic will be presented in~\cite{Lin:2026qvc}.

Ultimately, the shot-noise signal should hence help pin down the demographics of SMBH populations, especially when combined with measurements of the sky-averaged strain signal, local black hole mass determinations, and observations of distant quasars. 
Anisotropy detections will also help bound early-universe contributions to the GWB, which are expected to produce smaller anisotropies. On the other hand, the absence of detectable anisotropies might provide support for early-universe scenarios.
The SMBHB shot-noise anisotropy signal considered here should be accompanied by frequency-to-frequency scatter in the mean characteristic strain signal \cite{Sato-Polito:2024lew,Lamb:2024gbh}, as both features reflect the importance of bright rare sources. This also further motivates targeted searches for individual resolvable sources in PTAs. 

In principle, the LSS anisotropy signal offers a further powerful handle on the properties of SMBHB populations. The LSS anisotropies encode interesting information regarding the redshift distribution and clustering bias of the GWB sources. 
An appealing way to go after this signal is to cross-correlate PTA anisotropy maps with spectroscopic galaxy surveys.  
However, such measurements will remain challenging in the near term, as PTAs are currently sensitive only to large angular scales, while the small and subtle LSS fluctuations remain buried in the measurement noise. Nevertheless, continued PTA improvements promise to steadily expand the accessible anisotropy parameter space and help reveal the astrophysical origins of the GWB.

In future work, it may be interesting to compare our analytic calculations with simulations of GWB anisotropies. The simulations can be used to cross-check some of the simplifying approximations used here, such as the factorization adopted for the merger rate distribution (Eq.~\eqref{eq:dn}), the neglect of early merger events in the build-up of the present-day SMBH population, and the assumption of circular orbits. Other possible research directions include using our modeling framework for anisotropy measurement forecasts and to inject mock signals into end-to-end data pipelines to test signal recovery for different model parameters (e.g., \cite{Lemke:2024cdu,Konstandin:2025ifn,Schult:2025xpc,Petrov:2025afk}).

\acknowledgments
We thank 
Emily Liepold,
Stephen Taylor,
Wayne Hu, 
Ken Olum,
and
Gabriela Sato-Polito
for helpful discussions.
M-X.L. is supported by funds partially provided by the Canadian Institute for Theoretical Astrophysics (CITA) National Fellowship and funds provided by the Center for Particle Cosmology.
C-P.M. acknowledges the support of NSF AST-2307719.

\bibliography{ref}

\appendix
\section{Estimates for the SMBHB Clustering Bias Factors}
Here we discuss the simplifying assumption adopted in our model of a redshift-independent effective SMBHB bias factor. In a more detailed treatment of the clustering bias, one might connect the merging SMBHB populations to their host dark matter halos at each redshift of interest, anchoring the calculation to our theoretical understanding of halo bias. 

Here, we briefly consider a more detailed treatment along these lines, confining our
scope to a simple estimate. First, we suppose that 
$b(M_{\rm BH},q,z)$ (Eq.~\eqref{eq:beff}) is independent of $q$. Next, we assume that there is a correlation between the mass of a black hole and the mass of the dark matter halo it resides in. Specifically, we assume that the median black hole mass scales with its host dark matter halo mass, $M_{\mathrm{h}}$, as $M_{\rm BH} \propto M_{\mathrm{h}}^{\gamma}$.
We consider a superlinear scaling case with $\gamma=5/3$ (equivalently, $M_{\mathrm{h}} \propto M^{3/5}_{\rm BH}$) and a linear case with $\gamma=1$ for contrast.
A $\gamma \sim 5/3$ scaling might be expected in feedback-regulated scenarios for black hole growth \cite{Silk:1997xw,Wyithe:2003kc}, although perhaps with large scatter as black hole masses correlate more tightly with galaxy properties than with halo mass.
This scaling also receives some observational support: recent galaxy-galaxy lensing measurements around AGN with black hole mass estimates provide direct constraints on the $M_{\rm BH}$--$M_{\mathrm{h}}$ relation, and are consistent with the $M_{\rm BH} \propto M^{5/3}_{\mathrm{h}}$ dependence \cite{Li24}. Given that current uncertainties in the black hole and halo mass estimates are still fairly large, we explore $\gamma=1$ for comparison.

In the context of this model, the black hole mass function can be expressed as:
\begin{equation}
    \frac{dn}{dM_{\rm BH}} = \int dM_{\mathrm{h}} \frac{dn}{dM_{\mathrm{h}}} \frac{dP(M_{\rm BH}|M_{\mathrm{h}})}{dM_{\rm BH}},
\end{equation}
where $dn/dM_{\mathrm{h}}$ is the dark matter halo mass function, and $dP(M_{\rm BH}|M_{\mathrm{h}})/dM_{\rm BH}$ is the probability, per unit mass, that a black hole of mass $M_{\rm BH}$ resides in a halo of mass $M_{\mathrm{h}}$. We suppose that this follows a lognormal distribution, with the median black hole mass scaling as $M^{\gamma}_{\mathrm{h}}$, and that the scatter is independent of halo mass. In this case, the effective bias of Eq.~\eqref{eq:beff} may be written simply in terms of the halo mass function and halo bias as:
\begin{equation}
\label{eq:beffz_model}
    \langle{b_{\rm BH}(z)\rangle} \approx \frac{\displaystyle\int_{M_{\rm h,min}}^{M_{\rm h,max}} \! dM_{\mathrm{h}} \, \frac{dn}{dM_{\mathrm{h}}}(M_{\mathrm{h}}, z)\, M_{\mathrm{h}}^{5 \gamma/3} \, b(M_{\mathrm{h}}, z)}{\displaystyle\int_{M_{\rm h,min}}^{M_{\rm h,max}} \! dM_{\mathrm{h}} \, \frac{dn}{dM_{\mathrm{h}}}(M_{\mathrm{h}}, z)\, M_{\mathrm{h}}^{5\gamma/3}},
\end{equation}
where we have carried out the integrals over black hole mass in Eq.~\eqref{eq:beff}, effectively replacing them with integrals over halo mass, while noting that the lognormal scatter cancels out in the numerator and the denominator. The $M^{5\gamma/3}_{\mathrm{h}}$ scaling reflects the combination of the $\propto M_{\rm BH}^{5/3}$-weighting from the black hole mass dependence of the strain amplitude signal, along with the trend of black hole mass with halo mass, $\propto M^{\gamma}_{\mathrm{h}}$. Especially in the $\gamma=5/3$ case, the strong weighting with halo mass ($\propto M^{25/9}_\mathrm{h}$) implies that
the effective clustering bias at a given redshift is set primarily by the clustering of the most massive dark matter halos which host SMBHBs. For simplicity, we introduce mass cutoffs in the halo mass integral above. The 
results are insensitive to the lower mass cutoff adopted here, but -- at the low redshift end -- they do depend on the maximum mass cutoff included above. 

We calculate $\langle b_{\rm BH}(z) \rangle$ according to the above equation using the halo mass function model of \cite{Tinker:2008ff} and the halo bias model of \cite{Tinker10}. Although $\langle b_{\rm BH}(z) \rangle$ is a relatively strong function of $z$ in this model, it does not increase strongly enough with redshift to overcome the exponential suppression from the $p_z(z)$ factor in the redshift distribution (e.g., Eqs.~\eqref{eq:cl_exact}-\eqref{eq:lm_params}). The upshot of this is that this more elaborate model gives results that are quite similar to the simpler $\langle b_{\rm BH}(z) \rangle \propto {\rm constant}$ model. For example, adopting the above model for
$\langle b_{\rm BH}(z) \rangle$ with $\gamma=5/3$ and $M_{\rm h,max} = 10^{15.5} M_\odot$ in Eq.~\eqref{eq:cl_exact} gives angular power spectrum results that differ by less than $15\%$ from our $\langle b_{\rm BH} \rangle =5$ case. Effectively, the $p_z(z)$ models weight
low redshifts so strongly that the angular power spectrum results are only sensitive to $\langle b_{\rm BH}(z) \rangle$ close
to the peak redshift (which is not so far from $z\sim 0$, Fig.~\ref{fig:Wz}). In the simple model of Eq.~\eqref{eq:beffz_model}, the clustering bias near the peak redshift of $z \sim 0.5$ varies from $\langle b_{\rm BH}(z=0.5) \rangle = 1.3-5.6$ across the range $M_{\rm h,max} \sim 10^{13}-10^{16} M_\odot$. In the $\gamma=1$ case, the corresponding range is somewhat smaller,
$\langle b_{\rm BH}(z=0.5) \rangle = 1.1-3.0$.

It is also useful to compare these bias factors with those of luminous red galaxies (LRGs): as massive (stellar masses of $M_* \gtrsim 10^{11} M_\odot$), early-type galaxies they may be broadly representative of the SMBH host galaxies of interest here (with $M_{\rm BH} \sim 10^8-10^{10} M_\odot$). Recent DESI analyses find that the mean LRG host dark matter halos at $0.4 < z < 0.6$ (around our peak redshift) is $M_{\mathrm{h}} = 10^{13.4} M_\odot$, with a mean linear bias factor of $b=1.93$ \cite{Yuan:2023ezi}. The effective strain-weighted bias of Eq.~\eqref{eq:beff} may be larger than this number if the merging high-mass SMBHBs preferentially reside in especially massive and highly clustered LRGs.  
Hence, we believe our $\langle b_{\rm BH} \rangle = 1$, $5$ calculations bracket the likely range in effective clustering bias.

\end{document}